\documentclass[12pt,preprint]{aastex}

\usepackage{emulateapj5}
\usepackage{epsfig}
                                                                                
\input psfig.sty

\def\vkm{km s$^{-1}$}
\def\vkme{\textrm{km s}^{-1}}
\def\degree{$^\circ$}
\def\arcs#1{$#1''$}
\def\solarmass{$M_\odot$}
\def\solarmasse{M_\odot}

\def\Jyb{Jy beam$^{-1}$}
\def\mJyb{mJy beam$^{-1}$}
\def\Jybk{Jy beam$^{-1}$ km s$^{-1}$}
\def\tlabel#1{(\textit{#1})}

\def\cmc{cm$^{-3}$}
\def\micron{$\mu$m}
                                                                                
\def\ra#1#2#3#4{#1^\mathrm{h} #2^\mathrm{m} #3^\mathrm{s}_{^.} #4}
\def\dec#1#2#3{#1\degr #2\arcmin #3\arcsec}

\def\mH{m_\textrm{\scriptsize H}}
\def\Ro{R_\textrm{\scriptsize 0}}
\def\Rin{R_\textrm{\scriptsize in}}
\def\Rout{R_\textrm{\scriptsize out}}
\def\To{T_\textrm{\scriptsize 0}}
\def\no{n_\textrm{\scriptsize 0}}

\def\vphio{v_{\phi\textrm{\scriptsize 0}}}

\def\H2{H$_2$}
\def\N2HP{N$_2$H$^+$}
\def\HCOP{HCO$^+$}
\def\cCO{C$^{18}$O}
\def\cCOa{C$^{18}$O J=1-0}
\def\aCO{$^{12}$CO}
\def\bCO{$^{13}$CO}
\def\NH3{NH$_3$}

\def\putfig#1#2#3{\epsfig{scale=#1,angle=#2,figure=#3}}
\def\leftblank#1{}

\begin{document}

\title{Infall and Outflow around the HH 212 protostellar system}
\author{Chin-Fei Lee}
\affil{Harvard-Smithsonian Center for Astrophysics, SMA, 645, N. A'ohoku
Place, Hilo, HI 96720}
\email{E-mail:cflee@cfa.harvard.edu}
\author{Paul T.P. Ho, Henrik Beuther, Tyler L. Bourke, Qizhou Zhang}
\affil{Harvard-Smithsonian Center for Astrophysics, SMA, 60 Garden Street,
Cambridge, MA 02138}
\author{Naomi Hirano, Hsien Shang}
\affil{Academia Sinica Institute of Astronomy and Astrophysics, P.O. Box
23-141, Taipei 106, Taiwan}

\begin{abstract}
HH 212 is a highly collimated jet discovered in \H2{}
powered by a young Class 0
source, IRAS 05413-0104, in the L1630 cloud of Orion.
We have mapped around it in 1.33 mm continuum, 
\aCO{} ($J=2-1$), \bCO{} ($J=2-1$), \cCO{}
($J=2-1$), and SO ($J_K = 6_5-5_4$) emission
at $\sim$ \arcs{2.5} resolution with the Submillimeter Array.
A dust core is seen in the continuum around the source.
A flattened envelope is seen in \cCO{} around the source in the equator
perpendicular to the jet axis, with its 
inner part seen in \bCO{}.
The structure and kinematics of the envelope can be roughly 
reproduced by a simple edge-on disk model with both infall and rotation. 
In this model, the density of the disk is assumed to have a 
power-law index of $p=-1.5$ or $-2$, as found in other low-mass envelopes.
The envelope seems dynamically infalling toward the source
with slow rotation because the kinematics is found to be roughly consistent 
with a free fall toward the source plus a rotation of 
a constant specific angular momentum.
A \aCO{} outflow is seen surrounding the \H2{} jet, with a 
narrow waist around the source.
Jetlike structures are also seen in \aCO{} near the source
aligned with the \H2{} jet at high velocities.
The morphological relationship between the \H2{} jet and the
\aCO{} outflow, and the kinematics of the \aCO{} outflow along the jet axis
are both consistent with those seen in a jet-driven bow shock model.
SO emission is seen around the source 
and the \H2{} knotty shocks in the south, tracing shocked emission 
around them.
\end{abstract}


\section{Introduction}

Despite recent progress, the detailed physical processes in the early
stages of low-mass star formation are still uncertain. 
Currently, the most detailed and successful model of low-mass star formation
is the four-stage model proposed in the late 80's by \citet{Shu1987}. In the
first stage, slowly rotating molecular cloud cores form within a molecular
cloud as magnetic and turbulent support is lost through ambipolar diffusion.
In the second stage, a molecular cloud core becomes dynamically unstable and
collapses from the inside-out, forming a protostar surrounded by an
accretion disk deeply embedded within an infalling envelope of dust and gas.
In the third stage, as the protostar continues accreting mass through the
accretion disk, a stellar wind breaks out along the rotational axis of the
system, creating a bipolar outflow. In the fourth stage, the infall and
accretion terminate, leaving a newly formed star surrounded by a
circumstellar disk. In this paper, we investigate the protostellar collapse
and outflow in the second and third stages, with the Submillimeter Array
(SMA)\footnote{
The Submillimeter Array is a joint project between the Smithsonian
Astrophysical Observatory and the Academia Sinica Institute of Astronomy and
Astrophysics, and is funded by the Smithsonian Institution and the Academia
Sinica.} 
observations around a young protostar, IRAS 05413-0104, 
and its remarkable jet HH 212.

IRAS 05413-0104 is a cold, low-luminosity ($\sim$ 14 $L_\odot$), 
low-mass source \citep{Zinnecker1992}, located
at a distance of 460 pc in the L1630 cloud of Orion.
It is surrounded by a cold ($\sim$ 14 K), flattened 
(with an aspect ratio of 2:1)
rotating \NH3{} envelope with a diameter of 12,000 AU \citep{Wiseman2001}.
It is detected also at 1.1 mm \citep{Zinnecker1992} and 1.3 mm
\citep{Chini1997}, with a high
ratio of millimeter-wave luminosity to bolometric luminosity.
It is thus classified as a Class 0 source, which
has much material to be accreted still from the surrounding 
molecular gas envelope \citep{Andre2000}.

HH 212 is a remarkable jet with a length of $\sim$ 240 arcsecs ($\sim$ 0.6
pc) powered by the IRAS source.
It was discovered
in the rotational-vibrational $v=1-0$ S(1) emission line of \H2{} 
at $\lambda=2.122$ \micron{} \citep{Zinnecker1998}. 
It is highly collimated and symmetric
with matched pairs of shock knots (knotty shocks) 
and bow shocks on either side of the IRAS source.
Strong water masers are seen moving 
at $\sim$ 60 \vkm{} along the jet axis near the IRAS 
source \citep{Claussen1998}.
Shocked SiO emission is seen
along the jet axis \citep{Chapman2002} and around the source
\citep{Gibb2004}.
CO outflow is also seen surrounding the \H2{} jet \citep{Lee2000}.
Deep observations in \H2{} with the ESO Very Large Telescope (VLT)
show a pair of diffuse nebulae near the bases of the jet,
probably tracing the outflow cavity walls illuminated by the bright
knotty shocks around the source \citep{McCaughrean2002}.
Based on the relative magnitude of the
proper motions and radial velocities of the water masers, the CO 
outflow is believed to lie within 5\degree{} of the plane of the sky
\citep{Claussen1998}, and is thus an excellent system to 
investigate the transverse kinematics of outflow.

Here, we study in detail the circumstellar envelope around the IRAS source,
the molecular outflow surrounding the \H2{} jet, and the interaction
between them, with the SMA observations in 1.33 mm continuum, 
\aCO{} ($J=2-1$), \bCO{} ($J=2-1$), \cCO{}
($J=2-1$), and SO ($J_K = 6_5-5_4$) emission
at high angular and velocity resolutions.
It is currently believed that 1.33 mm continuum mainly traces dust core,
\aCO{} (particularly in interferometric
observations) mainly traces molecular outflow,
\cCO{} mainly traces envelope,
\bCO{} traces both molecular outflow and envelope, and
SO traces mainly shock interactions.

\section{Observations}

Observations in
1.33 mm continuum, 
\aCO{} ($J=2-1$), \bCO{} ($J=2-1$), \cCO{}
($J=2-1$), and SO ($J_K = 6_5-5_4$) emission
toward the IRAS source and the H$_2$ jet were made 
between 2004 November and 2005 March on top of Mauna Kea with the SMA
in the compact configuration (see Table \ref{tab:obs}).
Seven antennas were used in the array, giving baselines with projected
lengths ranging from 14.1 to 136 m, resulting in a synthesized beam
(with natural weighting) with a size of \arcs{2.8}$\times$\arcs{2.4} 
at 230 GHz.
With a shortest projected baseline of 14.1 m,
our observations were insensitive to structures more extended than 
\arcs{15} ($\sim$ 7000 AU) at the 10\% level.
The FWHP of the primary beam is \arcs{55}, and 9 pointings were used to map
the whole jet system. 
The digital correlator was set up with
a bandwidth of 104 MHz for each band (or chunk).
We used 512 spectral channels for \aCO{}, \cCO{}, and \bCO{} and 256 
spectral channels for SO,
resulting in a velocity resolution per channel
of $\sim$ 0.27 \vkm{} for \aCO{}, \cCO{}, and \bCO{}, and
0.55 \vkm{} for SO.
The 1.33 mm (or 225 GHz) continuum emission was also 
recorded with a total bandwidth of $\sim$ 4 GHz.
In our observation, a phase calibrator (Quasar 0423-013) 
was observed every 20 minutes to calibrate the phases of the source.
A planet (Uranus) was observed to calibrate the fluxes.
The data were calibrated with the MIR package adapted for the SMA.
The calibrated data were processed with the MIRIAD package.
The dirty maps that were produced from the calibrated data
and corrected for the primary beam attenuation
were deconvolved using the Steer clean method. 
The final channel maps were obtained by convolving the deconvolved maps
with a synthesized (Gaussian) beam fitted to the main lobe of the
dirty beam. 
The velocities of the channel maps are LSR.

\section{Results}

In the following, the IRAS source is assumed to have a 
refined position at $\alpha_{(2000)}=\ra{05}{43}{51}{408}$,
$\delta_{(2000)}=\dec{-01}{02}{53.13}$, which was
found at $\lambda= 3.5$ cm with the VLA at an
angular resolution of \arcs{0.3} \citep{Galvan2004}.
The systemic velocity in this region was assumed to be 1.8
\vkm{} in CO J=1-0 observations \citep{Lee2000}
and 1.6 \vkm{} in \NH3{} observations \citep{Wiseman2001}.
Here, it is assumed to be $1.7\pm0.1$ \vkm{}.
Throughout this paper, the observed velocity is 
the velocity with respect to this systemic velocity.
For comparison, our observations are always 
plotted with the \H2{} image of the jet 
recently obtained with the VLT \citep{McCaughrean2002}.
The axis of the \H2{} jet has a position angle of $\sim$ 23\degree{}.
The \H2{} jet is almost in the plane of the sky, with the 
blueshifted side to the north and the redshifted
side to the south of the source.

\subsection{1.33 mm Continuum Emission}

Continuum emission is detected around the source
at $\lambda=1.33$ mm with a flux of 0.11$\pm0.03$ Jy in between the diffuse
nebulae seen in \H2{} (Fig. \ref{fig:cont}).
This flux is similar to the peak flux in an \arcs{11} beam around the source
found by \citet{Chini1997} with the IRAM 30 telescope.
The emission is not resolved and shows a slight elongation along the jet axis.
Previously, continuum emission has been detected around the source
at $\lambda=1.1$ mm by the JCMT with a flux of 0.19 Jy 
\citep{Zinnecker1992}.
Continuum emission has also been detected in the infrared by the IRAS.
The fluxes corrected for extinction are
0.37, 19.9, and 61.9 Jy at $\lambda=25$, 60, and 100 $\mu$m, respectively. 
It was not detected at $\lambda=12$ $\mu$m.
The spectral energy distribution (SED, see Fig. \ref{fig:cont}b) is
consistent with a Class 0 source. It 
clearly suggests that the continuum emission at $\lambda=1.33$ mm is 
thermal emission from a dust core around the source.

The temperature of the dust core can be derived from the SED.
For simplicity, we assume a constant temperature for the dust core,
and the solid angle subtended by the dust core, $\Omega$, is
independent of frequency, $\nu$, so that the observed flux is given by
\begin{equation}
F_\nu = \Omega(1-e^{-\tau_\nu})B_\nu(T_{dust})
\end{equation}
where the optical depth
\begin{equation}
\tau_\nu = \kappa_\nu \sigma
\end{equation}
with $\kappa_\nu$ being
the mass opacity and $\sigma$ being the column density.
With \cite[see, e.g.,][]{Beckwith1990}
\begin{equation}
\kappa_\nu = 0.1 (\frac{\nu}{10^{12}\; \textrm{Hz}})^\beta \;\textrm{cm}^2
\;\textrm{g}^{-1}
\end{equation}
the optical depth can be simplified as
\begin{equation}
\tau_\nu = (\nu/\nu_o)^\beta
\end{equation}
with an optical depth of unity at $\nu_o$.
As can be seen from the SED, the emission are optically thin 
at $\lambda=1.1$ and 1.33 mm, and their flux ratio 
implies that $\beta\sim 1$.
The solid angle $\Omega$ is uncertain, 
likely $< 2.3\arcsec\times 1\arcsec$ (see Fig. \ref{fig:cont}a).
With $\Omega=1.2$ and 2.3 arcsec$^2$, the SED can be fitted with
$T_{dust}=47.6$ K and $\nu_o=3\times10^{12}$ Hz, 
and $T_{dust}=45.4$ K and  $\nu_o=6\times 10^{12}$ Hz, respectively (see
Fig. \ref{fig:cont}b).
Thus, $T_{dust}$ is assumed to be 46 K.
With this temperature and the optically thin emission at $\lambda=1.33$ mm,
the (gas $+$ dust) mass is estimated to be
\begin{eqnarray}
M \sim \frac{D^2 F_\nu}{B_\nu(T_{dust}) \kappa_\nu  } 
  \sim 0.08 M_\odot
\end{eqnarray}
where the distance to the source $D=460$ pc.


\subsection{\cCO{} ($J=2-1$) emission}

In \cCO{}, an envelope is seen around the source in 
the equator between the diffuse nebulae (Fig. \ref{fig:C18O}a).
Faint shells (2 $\sigma$ contours) are also seen opening to the north
and south surrounding the bases of the nebulae. 
The envelope is better seen in the blueshifted
emission, in which the shells are much fainter than the envelope
(Fig. \ref{fig:C18O}b).
It is extended across the source from the east to the
west with a position angle of $\sim 110$\degree{}$\pm10$\degree{}, 
perpendicular to the jet axis.
It is flattened with
an (deconvolved) FWHM extent of $\sim$ \arcs{4.3} (or 2000 AU at 460 pc) 
and an aspect ratio of $\sim$ 3:2.
In the redshifted emission, the envelope is less extended.
Since the redshifted emission around the bases of the
nebulae is as bright as that of the envelope,
the redshifted emission appears extended along the jet axis.

A cut along the major axis of the flattened envelope shows a PV structure
with the emission seen across the source with the redshifted and 
blueshifted peaks on either side of the source (Fig. \ref{fig:pvC18O}).
The blueshifted emission spreads across the source with a
peak at ($-0.5$ \vkm{}, $-$\arcs{0.45}) in the west, while
the redshifted emission spreads across the source with a peak
at (0.35 \vkm, \arcs{0.1}) in the east.
This PV structure is similar to that seen in \cCOa{} toward
an edge-on dynamically infalling envelope with rotation
around a young Class 0 source IRAS 04368+2557 in L1527 
\citep{Ohashi1997}.
It suggests that the material of the flattened envelope is also
infalling toward the source with rotation, with the redshifted 
emission from the nearside, while the blueshifted emission 
from the farside. 

\subsection{\bCO{} ($J=2-1$) emission}

In \bCO{}, the emission peaks at the source and extends to the bases
of the nebulae (Fig. \ref{fig:13CO}a). 
The emission can be decomposed into low-velocity and high-velocity components 
(see Fig. \ref{fig:pv13CO}b and the explanation in the following paragraph).
At low velocity, the emission forms a hourglass structure 
with a narrow waist around the source (Fig. \ref{fig:13CO}b), with
the blueshifted emission opening to the north and
the redshifted emission opening to the south around the bases of the nebulae
(Fig. \ref{fig:13CO}c).
Around the source, the blueshifted emission peaks 
a little to the west, while the redshifted emission shifts
to the east of the source and is weak along the major axis of the
flattened \cCO{} envelope.
At high velocity, the blueshifted emission peaks to the west while the
redshifted emission peaks to the east of the source (Fig. \ref{fig:13CO}d). 
The emission also
extends to the north and south along the jet axis.

A cut across the \bCO{} waist along the major axis of the
flattened \cCO{} envelope
shows that the \bCO{} emission has a larger
velocity range than that of the \cCO{} emission (Figs. \ref{fig:pv13CO}a
and \ref{fig:pv13CO}b).
The velocity range is thus decomposed into low-velocity 
and high-velocity components, 
with the range of the low-velocity component similar to that of 
the \cCO{} emission.
At low velocity, the blueshifted emission is seen across the source
along the major axis of the flattened envelope with a peak
at ($-0.3$ \vkm{}, \arcs{0.1}), and
a redshifted dip is seen around 0.4 \vkm{} across the source along the 
major axis of the flattened envelope. This dip may be
due to the absorption
of a cold material infalling toward the source \citep{Evans1999}, which is 
probably traced by the bright \cCO{} emission around that velocity (see Fig.
\ref{fig:pv13CO}b).
At high velocity, the redshifted and blueshifted emission 
are each confined on one side of
the source, as seen in Figure \ref{fig:13CO}d.
This high-velocity emission is unlikely to be from outflow,
with the redshifted and blueshifted emission 
on either side of the source.
This high-velocity emission may trace the inner part of the 
envelope, where the rotation starts to 
dominate over the infall motion and a rotationally supported
disk starts to form.

\subsection{\aCO{} ($J=2-1$) emission}

A \aCO{} outflow is seen surrounding the \H2{} jet, with a narrow
waist around the source (Fig. \ref{fig:12CO}). 
Near the source, limb-brightened 
outflow shells are seen opening to the north and south 
from the source surrounding the \H2{} jet.
The shells may extend further down along the jet axis, connecting to the
prominent \H2{} bow shocks SB4 and NB3, as seen in lower $J$ 
transition line of \aCO{} \citep{Lee2000}.
Deeper observations with shorter baselines are needed to
confirm this.
At high velocities, the \aCO{} emission is jetlike,
i.e., collimated and knotty, much like the \H2{} jet, 
located inside the cavity defined by the diffuse
nebulae and \aCO{} outflow shells (Fig. \ref{fig:12CO}b).
In the south, \aCO{} outflow shell is seen
coincident with the \H2{} bow shock SB4, probably tracing the ambient 
material swept up by the \H2{} bow shock.
Note that here the \aCO{} outflow shell is mainly seen on one side of the bow
shock, which is brighter in \H2{}.

Assuming an optically thin emission,
a temperature of 50 K, and a \aCO{} abundance of 8.5$\times10^{-5}$ with
respect to molecular hydrogen \citep{Frerking1982},
the \H2{} column density of the shells around the source
and the bow shock SB4 is found to be $\sim 10^{20}$ cm$^{-2}$.
Assuming that the shell thickness is $\sim$ \arcs{3} (see Fig. \ref{fig:12CO}),
the density is $\sim$ 5$\times10^3$ \cmc{}.
The peak column density along the jet axis is $\sim 5\times 10^{20}$
cm$^{-2}$. Assuming that the thickness is $\sim$ \arcs{2}, the density
is $\sim$ 4$\times10^4$ \cmc{}. 
Note that due to the absorption of the ambient cloud, that part of the
emission is optically thick, and that
part of the emission is resolved out by the interferometer,
the values presented here are lower limits of the true values.

This outflow, which is almost in the plane of the sky, is one of the best
candidates to study the outflow kinematics transverse to the jet axis.
A cut along the jet axis shows a series of convex spur PV structures
on both redshifted and blueshifted sides
with the highest velocity near the \H2{} bow tips and knots 
(Fig. \ref{fig:pv12CO}). 
Notice that the redshifted emission around $\sim 7$ \vkm{} 
merges with the foreground ambient cloud emission in velocity 
and thus is mostly resolved out from our observations.
These PV structures are similar to those seen
in lower $J$ transition line of \aCO{} \citep{Lee2000},
but with a higher velocity around the \H2{} knots and bow tips.
These PV structures suggest that the outflow
is accelerated by the shocks localized 
at the \H2{} bow tips and knots, so that the velocity of the outflow
decreases rapidly away from the tips,
similar to that seen in a pulsed jet simulation
\cite[see, e.g.,][]{Lee2001}.

\subsection{SO ($J_K = 6_5-5_4$) emission}

SO emission is detected around the source and the \H2{} knots in the
south, especially the knot SK4 (Fig. \ref{fig:SO}a). 
Around the source, the emission is not well resolved and 
is slightly elongated 
with a major axis about mid-way between the jet and envelope axes.
The blueshifted emission is to the north, while the
redshifted emission is to the south of the source (Fig. \ref{fig:SO}b), 
similar to that seen in the \H2{} jet, suggesting that
the SO emission may arise from interactions with the jet.
The peaks of the redshifted and blueshifted emission, however,
are a little off the jet axis, with the redshifted one shifted 
to the east and the blueshifted one shifted to the west, 
suggesting that the SO emission may also arise from the envelope.
Around the knot SK4, the redshifted emission is to the 
east while the blueshifted emission is to the west,
similar to that seen in the flattened \cCO{} envelope.

The SO emission around the source may consist of more than one component.
A cut along the jet axis shows two peaks at high velocity, 
one is blueshifted at
($-5.5$ \vkm{}, \arcs{0.64}) in the north and one is redshifted at
(6.1 \vkm{}, $-$\arcs{0.53}) in the south (Fig. \ref{fig:pvSO}a).
The rapid increase in the
velocity suggests that the SO emission at high velocity 
indeed arises from shock interactions with the jet.
However, the SO emission seen around the
systemic velocity may not be produced by the shock interactions, 
because of its low velocity.
A cut across the jet axis shows a rather symmetric PV structure
about the source (Fig. \ref{fig:pvSO}b) at low velocity.
The spatial extent and the velocity distribution (i.e., 
the redshifted emission is to the east and the blueshifted emission is
to the west) of this PV structure are similar to that 
of the flattened \cCO{} envelope, suggesting that the SO 
emission at low velocity may be associated with the flattened envelope.
There are also high blueshifted and redshifted emission at $\sim$ \arcs{3}
in the east.

A cut across SK4 shows that the redshifted emission is to the east, while
the blueshifted emission is to the west (Fig. \ref{fig:pvSO}c), 
as seen in Figure \ref{fig:SO}b.
The velocity extent seems to be similar to that around the source,
with the higher velocity near the jet axis.


\leftblank{
Masses are calculated assuming optically thin emission and that
the excitation temperature and the abundance relative to molecular hydrogen
are 20 K and 8.5$\times10^{-5}$ \citep{Frerking1982} for CO, 
10 K and 3$\times10^{-9}$ \citep{Girart2000}
for \HCOP{}, 10 K and $10^{-9}$ \citep{Linke1980}
for CS, and 7 K and 5$\times10^{-10}$ \citep{Caselli2002}
for \N2HP{}, respectively, integrating over the mapped area.
The total masses are 0.09, 0.10, 0.03, and 1.3 \solarmass{}, calculated
from the CO, \HCOP{}, CS, and \N2HP{} emission, respectively.
As for the ringlike structure itself, the mass is $\sim$ 0.5 $M_\odot$. 
Because of the absorption of the ambient cloud and the fact that 
part of the emission is resolved out by the interferometer, 
the values presented here are lower limits of the true values.
Note that the mass derived from the \N2HP{} emission is about 13 times 
larger than that derived from the CO and \HCOP{} emission, 
unless there is an abundance problem.
This may be partly because the CO and \HCOP emission
is resolved out by the interferometer around the systemic velocity (Fig.
\ref{fig:pvoutflow})
and partly because the \N2HP{} emission
indeed traces the dense envelope and ambient material.
Comparing to the single-dish observations, we find that
the CO flux within $\sim$ 1 \vkm{} of the systemic velocity 
is mostly resolved out by the interferometer.
The \HCOP{} and CS flux are also resolved out around the systemic 
velocity.}

\subsection{Summary}

In this section, we summarize our observations with two composite figures,
Figures \ref{fig:RGB}a and \ref{fig:RGB}b.
From these figures, it is clear that 
(1) the dust core seen in continuum and the \bCO{} waist are surrounded 
by the flattened \cCO{} envelope in the equator, and thus
may both trace the inner part of the envelope;
(2) the \bCO{} emission extends to the north and south from the source,
surrounding the bases of the nebulae;
(3) the \aCO{} outflow
appears as limb-brightened shells surrounding the \H2{} jet, and
(4) the SO emission is detected around the source and the \H2{} knots 
in the south.

\section{Modeling of Circumstellar Envelope}\label{sec:model}

In the following, a simple edge-on disk model with both infall and rotation
is used to reproduce the structures and kinematics of the
flattened \cCO{} envelope and the \bCO{} waist seen around the source.
In this model, the disk is assumed to have a constant thickness of $H$,
an inner radius of $\Rin$, and an outer radius of $\Rout$.
The number density of molecular hydrogen is assumed to be given by 
(in Cylindrical coordinates)
\begin{equation}
n = \no (\frac{R}{\Ro})^p
\end{equation} 
where $\no$ is the density at a characteristic radius of $\Ro=460$ AU (i.e.,
\arcs{1}) and $p$ is a power-law index.
The temperature of the envelope is uncertain and
assumed to be given by
\begin{equation}
T =  \To (\frac{R}{\Ro})^q
\end{equation}
where  $\To$ is the temperature at $\Ro$ and $q$ is a power-law index.
The abundances of \cCO{} and \bCO{} relative to molecular
hydrogen are assumed to be constant and given by 
$1.7\times10^{-7}$ and $9.5 \times 10^{-7}$, respectively
\citep{Frerking1982}.

The detailed radial profiles of the infall velocity and the rotation velocity
can not be determined from our observations obtained at current
spatial and velocity resolutions.
For simplicity,
we assume a dynamical infall that results in a free fall 
for the infall velocity. Assuming that the mass of the envelope is
small compared with that of the source, we have
\begin{equation}
v_R  = \sqrt{\frac{2GM_*}{R}} \sim
0.76 \sqrt{\frac{M_*}{0.15 \solarmasse}\frac{\Ro}{R}}
\;\;\; \vkme
\end{equation}
where $M_*$ is the mass of the source.
The rotation seems differential with the velocity increasing toward the
source (see Figs. \ref{fig:pvC18O} and \ref{fig:pv13CO}).
However, the rotation is unlikely to be
Keplerian, because the rotation velocity, as found in the following, is 
much smaller than the infall velocity.
In a dynamically infalling envelope with slow rotation, specific angular
momentum of each gas element is considered to be conserved, until the infall
motion shifts to the centrifugally supported motion around the radius where
the rotation velocity is comparable to the infall velocity \cite[see,
e.g.,][]{Nakamura2000}.
Thus, the rotation velocity is assumed to be given by
\begin{equation}
v_\phi =  \vphio \frac{\Ro}{R}
\end{equation}
where $\vphio$ is the rotation velocity at $\Ro$ and it depends on the
initial angular momentum at the outer radius.
In the model calculations, radiative transfer is used to calculate the emission,
with an assumption of local thermal equilibrium. For simplicity,
the line width is assumed to be given by the thermal line width only.
The line width due to turbulence is not included in our model.
The channel maps of the emission derived from the model are
convolved with the observed beams and velocity resolutions,
and then used to make the integrated maps and PV diagrams.

There are nine parameters in this model: $H$, $\Rin$, $\Rout$,
$\no$, $p$, $\To$, $q$, $M_*$, and $\vphio$. 
To narrow our search, only two promising values of $p$ are considered:
(1) $p=-1.5$, as found in many theoretical infalling models
\cite[see, e.g.,][]{Nakamura2000} and \HCOP{} 
observations around low-mass envelopes \citep{Hogerheijde2001},
and (2) $p=-2.0$, as found in 850 and 450 \micron{} continuum observations
of low-mass envelopes \citep{Shirley2000}.
We find that these values of $p$ 
can both result in a reasonable fit to our observations too
(see Fig. \ref{fig:fpvenv}). 
In addition, the values of other parameters
do not depend much on these values of $p$.
In our fits, we have $H \sim 550$ AU ($\sim$ \arcs{1.2}), 
$\Rin \sim 115$ AU ($\sim$ \arcs{0.25}), 
$\Rout \sim 2300$ AU ($\sim$ \arcs{5}), $M_* \sim 0.15$\solarmass{}, 
$\vphio \sim 0.3$ \vkm{}, and $\no \sim 4 \times 10^ 6$ \cmc{}.
$\To$ and $q$, however, are required to be different between \cCO{} and
\bCO{} in order to fit the observations.
We have $\To \sim 35$ K and $q\sim -0.4$ for \cCO{},
and $\To \sim 23$ K and $q\sim -1$ for \bCO{}.
With these values of parameters, our model can roughly reproduce: 
(1) the observed structures of the flattened \cCO{} envelope and the 
\bCO{} waist seen around the source;
(2) the observed PV structures with the emission seen across the source with 
the redshifted and blueshifted peaks on either side of the source; and
(3) the observed redshifted dip in \bCO{}.
Note that in the observations,
the emission extending to the north and south is
likely from the envelope material affected by the outflow shells 
(see \S \ref{sec:env}) and thus is not included in our model.

The mass of the flattened envelope, which is given by
\begin{eqnarray}
M_e &=& 2 \mH \no H \int_{\Rin}^{\Rout} (\frac{R}{\Ro})^p 2 \pi R dR
\end{eqnarray} 
is $\sim 0.028$ and 0.025 \solarmass{} for $p=-1.5$ and $-2$, respectively.
This mass, however, is a factor of 3 lower than that of the dust core.
Note that the mass derived for the dust core could have an uncertainty
with a factor of 5 due to the large uncertainty in the mass opacity 
\citep{Beckwith1990}.
The infall rate, which is given by
\begin{eqnarray}
\dot{M}(R) &= &2 \pi R H n |v_R| 2 \mH
\end{eqnarray}
is $\sim 6 \times 10 ^{-6} M_\odot\; \textrm{yr}^{-1}$ at $\Ro=460$ AU.

\section{Discussion}

HH 212 is a highly collimated jet powered by a young Class 0
source, IRAS 05413-0104.
We have mapped the jet and its source in 1.33 mm continuum, 
\aCO, \bCO{}, \cCO{}, and SO.
In the following, we discuss our results in details.

\subsection{Circumstellar Envelope} \label{sec:env}

A flattened envelope is seen in \cCO{} around the source in the equator
perpendicular to the jet axis, 
with its inner part seen in \bCO{}.
The structure and kinematics of the envelope can be roughly 
reproduced by a simple edge-on disk model with both infall and rotation. 
In this model, the density of the disk is assumed to have a 
power-law index of $p=-1.5$ or $-2$, as found in other
low-mass envelopes \citep{Shirley2000,Hogerheijde2001}.
The temperature of \cCO{} has a power-law index of $q\sim -0.4$,
as found in other low-mass envelopes \citep{Shirley2000,Hogerheijde2001}.
However, in order to reproduce the compact waist
and the redshifted dip in \bCO{}, the temperature of \bCO{} is required to
decrease with $q\sim -1$, which is faster than that of \cCO{}.
Notice that in our model, the redshifted dip is assumed to be produced by the 
absorption of a cold material at the outer part of the envelope.
However, the dip may be partly due to missing flux in our 
interferometric observations \citep{Gueth1997}.
Further observations and modeling 
(e.g., with flared disk model, turbulence, and missing flux consideration)
are needed to improve this.

The envelope seems dynamically infalling toward the source
with slow rotation because the kinematics is found to be roughly consistent 
with a free fall toward the source plus a rotation of 
a constant specific angular momentum.
If this is the case, the source has a mass of $M_*\sim 0.15$ \solarmass{} 
and the infall rate is
$\sim 6 \times 10 ^{-6} M_\odot\; \textrm{yr}^{-1}$ at $\Ro=460$ AU,
similar to those found in other flattened infalling envelopes with 
slow rotation around low-mass protostars \citep{Ohashi1997,Momose1998}.
Assuming that the accretion rate is 
constant in the past and given by the infall rate, 
the accretion time will be $M_*/\dot{M}\sim 3 \times 10^4$ yr, as expected
for a Class 0 source.
Notice that in theoretical models \cite[e.g.,][]{Shu2000}, 
this accretion rate corresponds to an isothermal sound speed of 0.3 \vkm, 
or equivalently a temperature of 25 K, which is consistent with our
simple model.
This dynamically infalling envelope with
rotation is expected to form a rotationally supported disk at 
$R \sim 74$ AU ($\sim$ \arcs{0.16}), at 
which the rotation velocity is comparable to the infall velocity
\citep{Hayashi1993,Lin1994}.
The \bCO{} emission seen at high velocity may arise
from this rotationally supported disk, with the redshifted emission and
the blueshifted emission on either side of the source (Fig. \ref{fig:13CO}d).


\leftblank{
This Keplerian rotation
requires a mass of 0.1 solar mass, similar to that derived from
the continuum emission. The disk mass is estimated to be ~ 0.02 solar
mass, smaller than the mass.
The detailed velocity structure is unclear. Assuming a Keplerian rotation
\begin{equation}
v_r = \sqrt{\frac{GM}{r}} = 0.5 \sqrt {\frac{0.8"}{r}}
\end{equation}
with $r_o=0.8"$, $v_{ro}=0.8"$.
Thus, the mass associated with the center star is
\begin{equation}
M = r_o v_{ro}^2/G = 0.1 \solarmasse
\end{equation}
}

A flattened \NH3{} envelope has been seen rotating around the source
in the same direction \citep{Wiseman2001}.
It has a characteristic radius of $\sim$ 3500 AU ($\sim$ \arcs{8}) 
and can be considered as the extension of the flattened \cCO{} envelope 
into the surrounding. At that radius, 
the infall velocity and rotation velocity are expected to be 
$\sim$ 0.27 and 0.04 \vkm{}, respectively in our simple model.
This rotation velocity is consistent with that observed in the \NH3{}
envelope, suggesting that the rotation of the \cCO{} envelope is connected
to that of the \NH3{} envelope, and that the angular momentum is carried 
inward into the \cCO{} envelope from the \NH3{} envelope. However,
no infall velocity has been seen in the \NH3{} envelope.
In theoretical models \cite[e.g.,][]{Shu1977},
the infall radius is given by the isothermal sound speed times the
accretion time. Assuming a temperature of 25 K,
the isothermal sound speed is $\sim$ 0.3 \vkm{}, resulting in
an infall radius of
$\sim$ 2000 AU, similar to the outer radius of the \cCO{}
envelope. Thus, it is possible that no infall motion is seen in the \NH3{}
envelope.


\leftblank{
From the PV diagrams of both \bCO{} and \cCO{}, it is clear that
the low-velocity material is about the same size in
both blue \bCO{} and red \cCO{}, with a radius of about \arcs{2}. 
Thus, \bCO{} probes two components: an infall
envelope (as \cCO{}) and a rotating inner disk or envelope.
The inner envelope is not seen in \cCO{} likely due to sensitivity.
The redshifted dip is probably due the absorption of a cold 
infalling material at the outer part. However, the observed size of
\bCO{} is smaller than that of \cCO{}, which means the \cCO{} extends
further out than the \bCO{}. If the temperature decreases outward, then
it is difficult to explain the \bCO{} redshifted dip.
}


The flattened envelope is not formed by rotation because 
it is not rotationally supported,
with the rotation velocity smaller than the infall velocity.
Could it result from an interaction with the CO outflow?
The flattened \NH3{} envelope is seen carved by the CO outflow,
forming a bowl structure around it \citep{Wiseman2001,Lee2000}.
In our observations, the shells seen in \cCO{} and \bCO{} around
the bases of the nebulae may trace the (swept-up) envelope material
pushed by the outflow shells toward the equator, suggesting that
the envelope around the source is also carved by the outflow.
Having said that,
the flattened envelope may not form by the outflow interaction,
because the shells are seen separated from the envelope.

Formation mechanisms of such a flattened infalling envelope
around a forming star have been proposed
with and without magnetic field.
\citet{Galli1993} examined the gravitational contraction of 
magnetized spherical cloud cores.
They showed that flattened infalling envelopes are formed as
the magnetic field impedes the contraction perpendicular to the field lines, 
which are aligned with the rotation axis.
\citet{Hartmann1996} showed that even in the absence of 
a magnetic field, initially sheetlike cloud
cores can also collapse to form flattened envelopes.
Recently, \citet{Nakamura2000} also showed that 
flattened infalling envelopes 
are a natural outcome of the gravitational
contraction of prolate cloud cores with slow rotation.
In this scenario, the flattened envelopes are predicted to have
$n\propto r^{-1.5}$, $v_r \propto r^{-0.5}$, and $v_\phi \propto r^{-1}$,
the same as our model.

\subsection{Molecular Outflow}
A \aCO{} outflow is seen surrounding the \H2{} jet, with a
narrow waist around the source.
The outflow seen here in \aCO{} $J=2-1$ is
similar to that seen in \aCO{} $J=1-0$,
but with higher $J$ transition line of \aCO{} tracing higher velocity
around the knots and bow tips.
The morphological relationship between the \H2{} jet and the
\aCO{} outflow, and the kinematics of the \aCO{} outflow along the jet axis
are both consistent with those seen in a jet-driven bow shock model
\cite[see, e.g., ][and reference therein]{Lee2001}.
In that model, a highly collimated jet is 
launched from around the source, propagating into the ambient medium.
A temporal variation in the jet velocity produces a chain of 
\H2{} knots propagating down along the jet axis.
Since these \H2{} knots are localized (ballistic) shocks with high thermal pressure,
they expands sideways, growing into \H2{} bow shocks 
with the shock velocity (and thus temperature)
decreasing rapidly away from the bow tips.
As these bow shocks propagate down along the jet axis, they
sweep up the ambient material into a thin outflow shell around the jet axis.
In our observations, the \aCO{} shells around the source
can be identified as the ambient material swept up by the \H2{} bow shocks.
In addition, \aCO{} emission also arises from around the \H2{} knots and bow shocks,
with the velocity decreasing rapidly away from the knots and bow tips.
Higher $J$ transition line of \aCO{} traces higher velocity around the knots and
bow tips likely because
it arises from regions of higher temperature and thus higher velocity 
closer to the knots and bow tips. 
As a result, the \aCO{} emission appears jetlike at high velocities
in high $J$ transition.
Located inside the outflow cavity, the jetlike \aCO{} emission is unlikely
from the entrained ambient gas. 
The jet is probably intrinsically molecular, 
such as that seen in HH 211 \citep{Gueth1999}.
Since the shell is the swept-up ambient material, the density of the
ambient material is expected to be lower than that of the shell. Thus,
the jet is an overdensed jet with the density higher than that 
of the ambient material.

In the south, \aCO{} emission is seen coincident with the \H2{} bow
shock SB4, indicating that the shocked material there must have cooled 
very fast from $\sim$ 2000 K (\H2{}) to a few 10 K (\aCO{}) because
of a significant mixing with the ambient material.
The transverse velocity there is also lower than that 
around other bow shocks (see Fig. \ref{fig:pv12CO}), 
probably because the bow shock SB4 is sharing momentum with 
the quiescent ambient material \citep[cf. Fig. 7 in][]{Lee2001}.
Therefore, the bow shock SB4 may trace the leading
bow shock where the head of the jet impacts on the ambient material.

\leftblank{
At $i=0$\degree{}, the PV diagram in the ballistic bow shock model
can be expressed as
$v_{obs} = [\beta c_s v_s^2 R_j^2/9]^{1/3} |z|^{-2/3} $,
where $|z|$ is the projected distance from the bow shock.
For an internal bow shock, this relationship becomes
$v_{obs}=[\beta c_s \triangle v (v_{si}-v_e) R_j^2/(18 \rho_e/\rho_j)]^{1/3}|z|^{-2/3}$,
where $v_{si}$ is the internal shock speed, $\triangle v$ is velocity jump
across the internal shock, $v_e$ and $\rho_e$ are the velocity and density
of the cocoon material \citep{Ostriker2001}.
}

\leftblank{
As mentioned,
the CO outflow is driven by the \H2{} bow shocks. Since the
\H2{} bow shocks are likely driven by an underlying highly collimated
jet, a simple jet-driven ballistic bow shock model is used to explain the
\H2{} bow-shock structure and the CO outflow kinematics. As can be seen,
this simple model can roughly reproduce the bowshock structures and CO
outflow kinematics for SB3 and SB4.
}

\leftblank{
In this model, the shock is localized at the tip of the bow. The shocked
material is ejected sideways away from the bow tip, forming a bowlike
structure as it moves down along the jet axis. Material is piling up at
the bow shock, sharing the momentum with the shocked material, so that the
transverse velocity decreases rapidly from the bow tip to the wing.
}

\subsection{SO emission}

SO emission is seen around the source 
and the \H2{} knotty shocks in the south, especially the knot SK4.
The emission around the source may arise from two components.
Like that around the knot SK4, the emission seen at high velocity is likely to be
from knotty shocks, which are expected to be
there following the \H2{} knotty shocks toward the source.
The shocks are not seen in \H2{} probably due to dust extinction.
However, the SO emission seen around the systemic velocity may arise
from the flattened envelope due to its low velocity.
Observations at high angular resolution are needed to confirm this.

SO emission has also been seen toward the CepA-East outflows,
tracing ambient material at 60-100 K and shocked gas at 70-180 K 
\citep{Codella2005}.
Assuming an optically thin emission, a temperature of 120 K, 
and a SO abundance of 10$^{-9}$ relative to molecular hydrogen 
\citep{Codella2005}, %
the mass is found to be $\sim$ 0.4 and 0.1 \solarmass{}, respectively, 
around the source and the knot SK4. 
However, these masses are too large to be from knotty shocks.
The abundance of SO must have been highly enhanced.
Note that \aCO{} emission is weak around the knot SK4 (see Fig.
\ref{fig:12CO}b), 
probably suggesting that \aCO{} there is destroyed 
by the shock interaction.
The detection of the SO emission is likely because of the abundance
enhancement of SO due to the shock-triggered release of various molecules
(e.g., H$_2$S/OCS/S) from dust mantles and the subsequent chemical reactions
in the warm gas. Since the abundance is enhanced only for a very short
period of time, the SO emission traces regions of recent shock activity 
\citep{Codella2005}.

Around the knot SK4, the redshifted emission is seen to the east
while the blueshifted emission is seen to the west, 
similar to that seen in the flattened \cCO{} envelope. 
A similar velocity structure has been seen around the
knot SK1 (see Fig. \ref{fig:SO} for its location)
in \H2{} and considered as a tentative evidence
that the knot SK1 is rotating \citep{Davis2000}. 
It was observed with 3 slits from east to west
with a separation of $\sim$ \arcs{0.45}.
The peak velocities with respect to the systemic
velocity were found to be +4.6, +2.9, and +2.3 \vkm{}, measured from east to west. 
Thus, assuming that the central velocity at the knot SK1 is $\sim 3$ \vkm{}, 
the redshifted emission is seen to the east while the blueshifted is seen
to the west. 
A similar velocity structure is seen here around the knot SK4 in SO
but with higher velocity. 
Does this suggest that the knot SK4 is also rotating
and that the jet carries away angular momentum from the infalling envelope?
The velocity here, however, is much larger than $\vphio$ (i.e., 0.3 \vkm{}) 
and thus may be too large to be from rotation. 
In addition, the shocked material there is expected to have sideways expansion. 
Further observations at higher angular resolution are needed to study it.

\leftblank{
Peak radial speeds are (+6.3, +4.6,+4.0) km/s
W.R.T. Vsys=1.7 km/s   (+4.6, +2.9,+2.1) km/s
Note the central slit is shifted by 0.1" to the west of the jet axis.
So jet central velocity is expected to be ~ 3. km/s
}

\leftblank{
(The scenario proposed by most models is that H2S is the main reservoir of
sulphur on grain mantles: once in the gas phase, H2S is used for a fast
production of SO and SO2 (e.g., Pineau des Forets et al. 1993, Charnley
1997). However, the lack of H2S features in ISO spectra (Gibb et al. 2000,
Boogert et al. 2000) which set upper limits on the iced H2S abundance around
protostars and the detection of OCS on grains (Palumbo et al. 1997) suggest
that the latter may be an important sulphur carries in the ices. This seem
supported by the observations in the envelopes of massive young stars
recently performed by van der Tak et al. (2003), which indicate for OCS
higher excitation temperatures than for H2S. An alternative hypothesis is
that the sulphur released from the dust mantles is mainly in atomic form
(Wakelam et al. 2004).)
}


\section{Conclusion}
We have mapped the 1.33 mm continuum,
\aCO{} ($J=2-1$), \bCO{} ($J=2-1$), \cCO{}
($J=2-1$), and SO ($J_K = 6_5-5_4$) emission
around a protostellar jet HH 212 and its central source, IRAS 05413-0104.
A dust core is seen in the continuum around the source,
with a temperature of $\sim$ 46 K and a mass of $\sim$ 0.08 \solarmass{}.
A flattened envelope is seen in \cCO{} around the source in the equator
perpendicular to the jet axis, with its inner part seen in
\bCO{}.
The structure and kinematics of the envelope can be roughly 
reproduced by a simple edge-on disk model with both infall and rotation. 
The flattened envelope seems dynamically infalling toward the source
with slow rotation because the kinematics is found to be roughly consistent 
with a free fall toward the source plus a rotation of 
a constant specific angular momentum.
A \aCO{} outflow is seen surrounding the \H2{} jet, with a 
narrow waist around the source. 
Jetlike structures are also seen in \aCO{} near the source
aligned with the \H2{} jet at high velocities.
The morphological relationship between the \H2{} jet and the
\aCO{} outflow, and the kinematics of the \aCO{} outflow along the jet axis
are both consistent with those seen in a jet-driven bow shock model.
SO emission is seen around the source 
and the \H2{} knotty shocks in the south, tracing shocked emission 
around them.

\acknowledgements
We thank the SMA staff for their efforts
in running and maintaining the array.
H.B. acknowledges financial support by the Emmy-Noether-Program of the
Deutsche Forschungsgemeinschaft (DFG, grant BE2578/1).



\begin{deluxetable}{lrrccc}
\tablecolumns{6}
\tabletypesize{\normalsize}
\tablecaption{Summary of Observations
\label{tab:obs}}
\tablewidth{0pt}
\tablehead{
\colhead{Line} & \colhead{Frequency}   &
\colhead{Beam Size}  & \colhead{Channel Width} & \colhead{Noise Level$^a$ $\sigma$} \\
               & \colhead{(GHz)}       & 
\colhead{(arcsec $\times$ arcsec)}   & \colhead{(\vkm{})} &\colhead{(\Jyb{})}  
}
\startdata
1.33   mm continuum            & 225.000000 &  2.6$\times$2.3 & ...   &3.5E-3  \\
\aCO{} $J=2-1$       & 230.537980 &  2.8$\times$2.3 &0.264  & 0.25  \\
\bCO{} $J=2-1$       & 220.398676 &  2.8$\times$2.4 &0.274  & 0.15  \\
\cCO{} $J=2-1$       & 219.560357 &  2.8$\times$2.4 &0.274  & 0.14  \\
 SO    $J_K=6_5-5_4$ & 219.949442 &  2.8$\times$2.4 &0.554  & 0.12  \\
\enddata
\tablenotetext{a}{RMS noise level per velocity channel}
\end{deluxetable}
\clearpage

\begin{figure} [!hbp]
\centering
\putfig{0.9}{270}{f1.ps}
\figcaption[]
{
\tlabel{a} 1.33 mm continuum contours plotted on top of the 
\H2{} image adopted from \citet{McCaughrean2002}. The contours go from 3 to 18 $\sigma$ with a step of 3
$\sigma$, where $\sigma = 3.5$ \mJyb{}. 
The asterisk indicates the IRAS source.
Note that the (0, 0) position here corresponds to 
$\alpha_{(2000)}=\ra{05}{43}{51}{30}$,
$\delta_{(2000)}=\dec{-01}{02}{53.0}$.
\tlabel{b} Simple fits to the SED of the continuum source with 
$\Omega=1.2$ (solid line) and 2.3 (dashed line) arcsec$^2$.
\label{fig:cont}}
\end{figure}

\begin{figure} [!hbp]
\centering
\putfig{0.85}{270}{f2.ps}
\figcaption[]
{\cCO{} emission plotted on top of the \H2{} image, 
with the asterisk indicating the IRAS source.
\tlabel{a} Total emission integrated over 2.2 \vkm{} from $-$1.1
\vkm{} to 1.1 \vkm{}. The contours go from 2 to 23 $\sigma$ with a step of 3
$\sigma$, where $\sigma = 0.11$ \Jyb{}.
\tlabel{b} Redshifted (red contours) and blueshifted 
(blue contours) emission, 
integrated over 1.1 \vkm{} on the redshifted and blueshifted side, 
respectively. The contours go from 
2 to 17 $\sigma$ with a step of 3 $\sigma$, where $\sigma = 0.08$ \Jybk{}.
\label{fig:C18O}}
\end{figure}

\begin{figure} [!hbp]
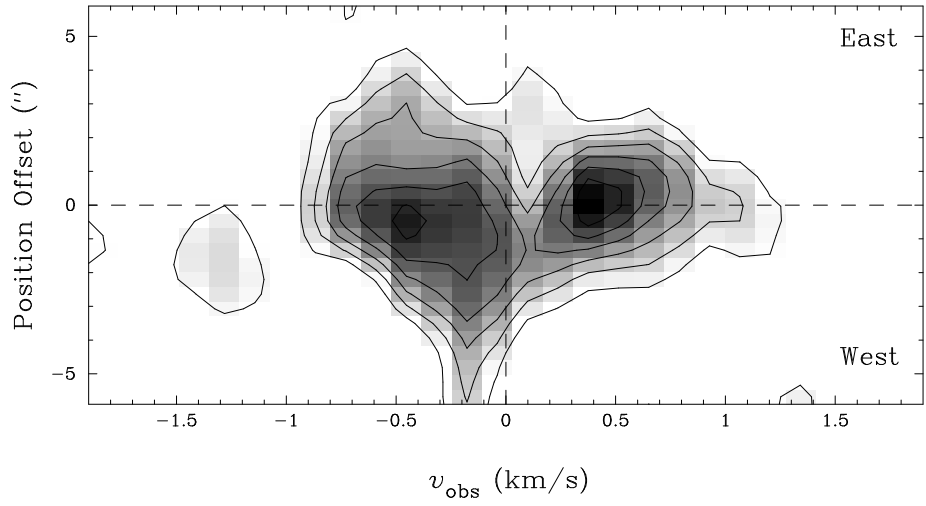

\centering
\putfig{0.7}{0}{f3.ps}
\figcaption[]
{PV diagram of \cCO{} emission cut along the major axis of
the flattened envelope.
\label{fig:pvC18O}
}
\end{figure}

\begin{figure} [!hbp]
\centering
\putfig{0.85}{270}{f4.ps}
\figcaption[]
{\bCO{} emission plotted on top of the \H2{} image, 
with the asterisk indicating the IRAS source. 
\tlabel{a} Total emission integrated over 3.4 \vkm{} from -1.7 \vkm{} to
1.7 \vkm{}. The contours go from 2 to 20 $\sigma$ with a step of 3 $\sigma$,
where $\sigma = 0.15$ \Jybk{}.
\tlabel{b} Low-velocity emission integrated from -0.7 to 0.7 \vkm{}.
The contours go from 2 to 17 $\sigma$ with a step of 3 $\sigma$,
where $\sigma = 0.09$ \Jybk{}.
\tlabel{c} Low-blueshifted (blue contours) and low-redshifted (red contours)
emission, integrated from -0.7 to 0 \vkm{} and 0 to 0.7 \vkm{},
respectively.
The contours go from 2 to 18 $\sigma$ with a step of 2 $\sigma$,
where $\sigma = 0.06$ \Jybk{}.
\tlabel{d} High-blueshifted (blue contours) and high-redshifted (red contours)
emission, integrated from -1.7 to -0.7 \vkm{} and 0.7 to 1.7 \vkm{},
respectively. The contours go from 2 to 14 $\sigma$ with a step of 2 $\sigma$,
where $\sigma = 0.08$ \Jybk{}.
\label{fig:13CO}
}
\end{figure}

\begin{figure} [!hbp]
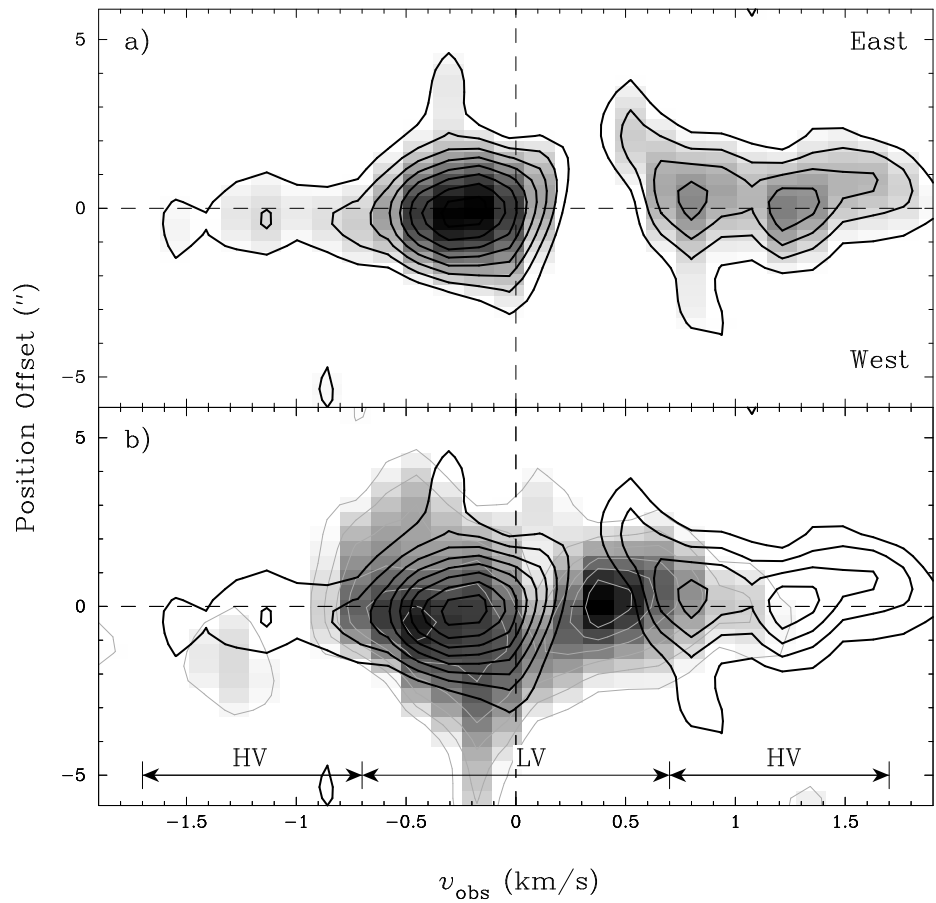

\centering
\putfig{0.7}{0}{f5.ps}
\figcaption[]
{PV diagrams of \cCO{} and \bCO{} emission cut along the major axis of
the flattened envelope. \tlabel{a} for \bCO{}. 
\tlabel{b} Overlay of the two. Here, LV denotes low velocity and
HV denotes high velocity.
\label{fig:pv13CO}
}
\end{figure}

\begin{figure}[!hbp]
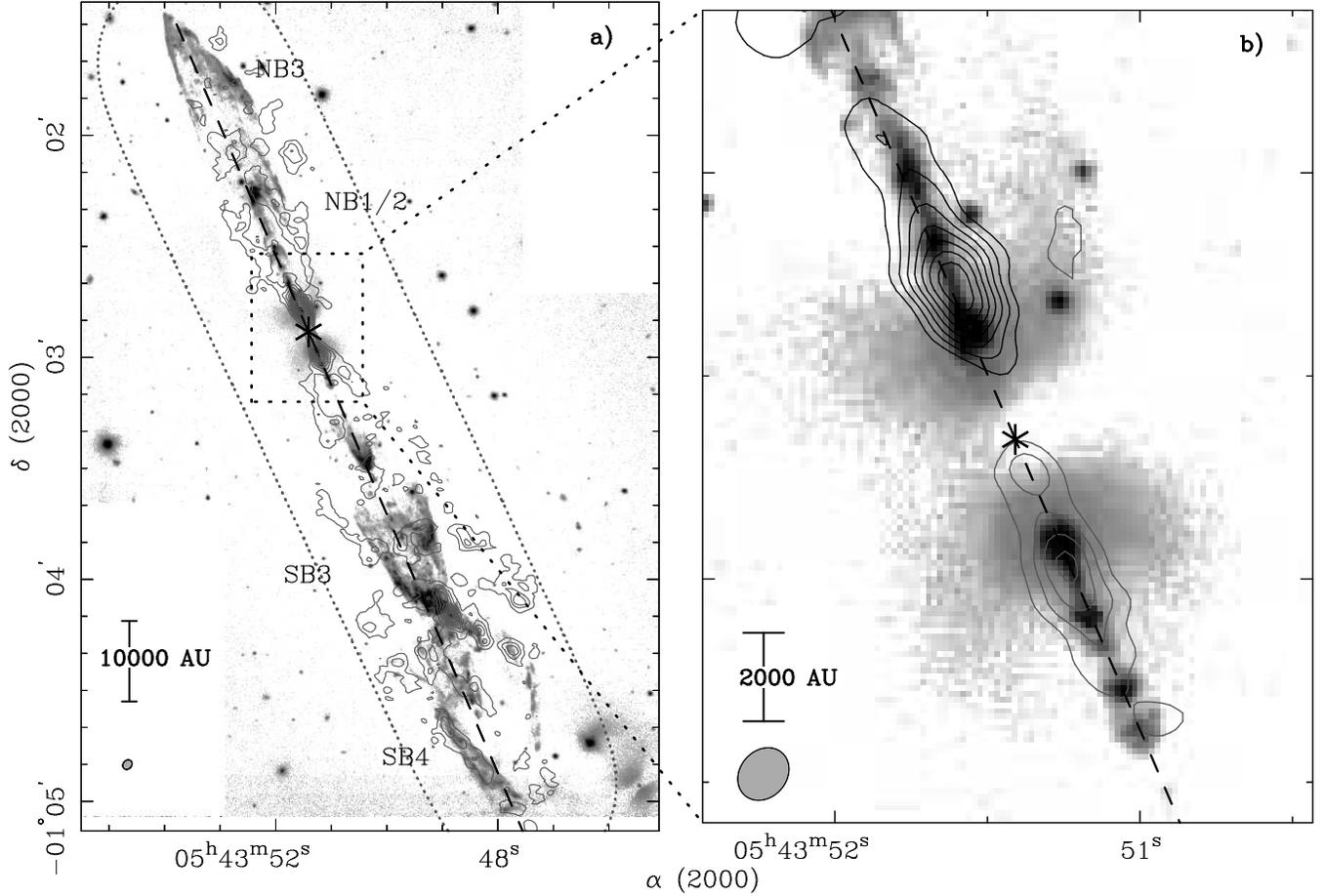

\centering
\putfig{0.7}{270}{f6bw.ps}
\figcaption[]{
\aCO{} emission plotted on top of the \H2{} image adopted from
\citet{McCaughrean2002},
with the asterisk indicating the IRAS source.
The black dotted line outlines the observed region in our observations.
\tlabel{a} Black contours: 
Total emission integrated over 32 \vkm{} from $-$16 to 16 \vkm{}.
The contours go from 
2 to 24 $\sigma$ with a step of 2 $\sigma$, where $\sigma = 0.68$ \Jybk{}.
\tlabel{b} Black contours: High blueshifted emission
integrated from $-$5 to $-$16 \vkm{}. The contours go from 
3 to 21 $\sigma$ with a step of 3 $\sigma$, where $\sigma = 0.42$ \Jybk{}.
Gray contours: High redshifted emission
integrated from 2 to 13 \vkm{}. The contours go from 
3 to 12 $\sigma$ with a step of 3 $\sigma$, where $\sigma = 0.42$ \Jybk{}.
Here, the velocity ranges are selected to show the
jetlike structures as much as possible, without the confusion from the
shell emission.
\label{fig:12CO}
}
\end{figure}

\begin{figure}[!hbp]
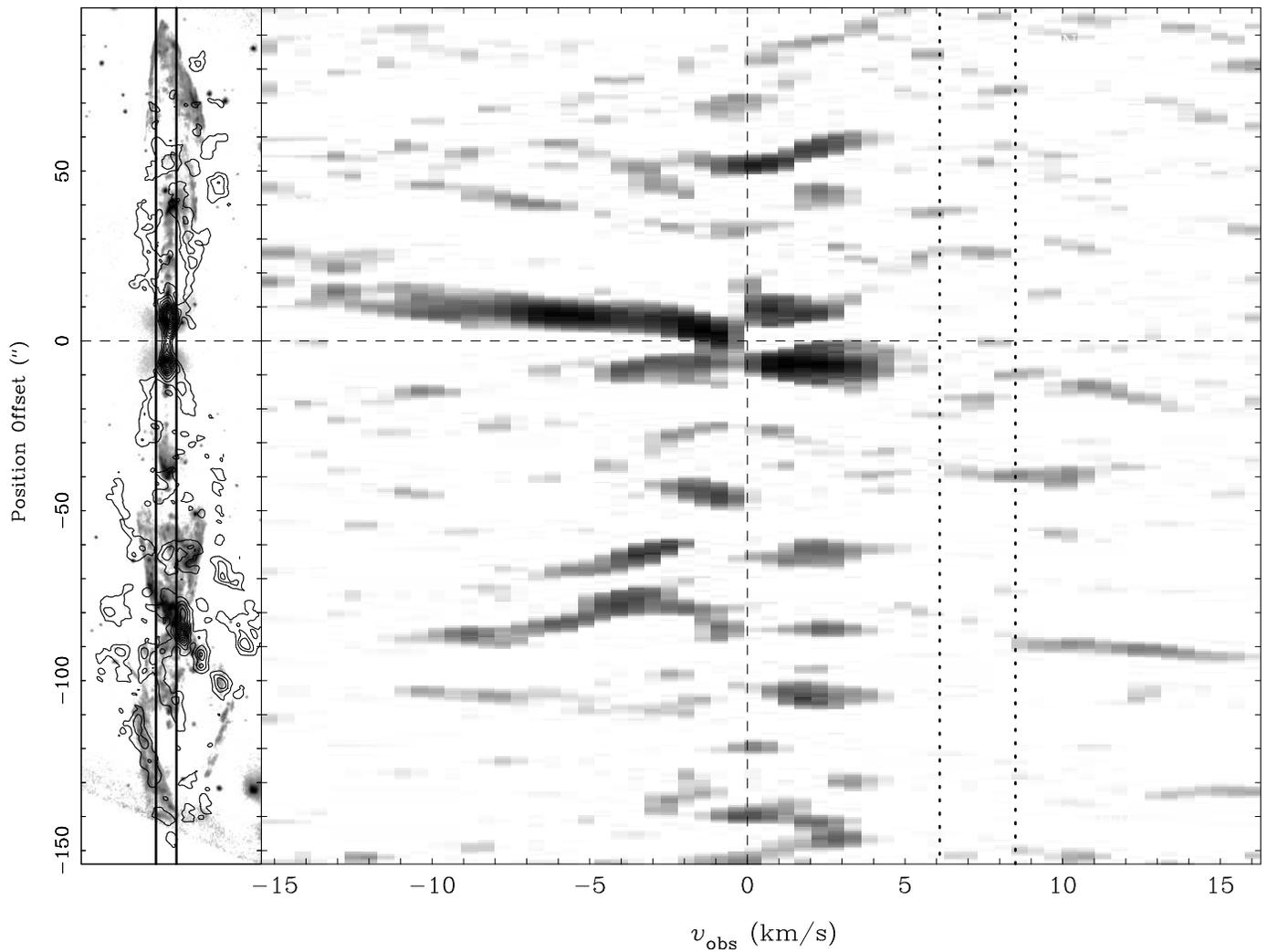

\centering
\putfig{0.7}{270}{f7.ps}
\figcaption[]{
{\it Left}: The \H2{} jet and CO outflow, with the outflow axis rotated to be
aligned with the y-axis, The two vertical lines indicate the width of the
cut for the position-velocity diagram. {\it Right}: 
Position-velocity diagram of the CO outflow cut along the
outflow axis. The two dotted lines around 7 \vkm{} indicate the 
velocity range of a foreground ambient cloud seen by the FCRAO single-dish
telescope \citep{Lee2000}.
\label{fig:pv12CO}
}
\end{figure}

\begin{figure}[!hbp]
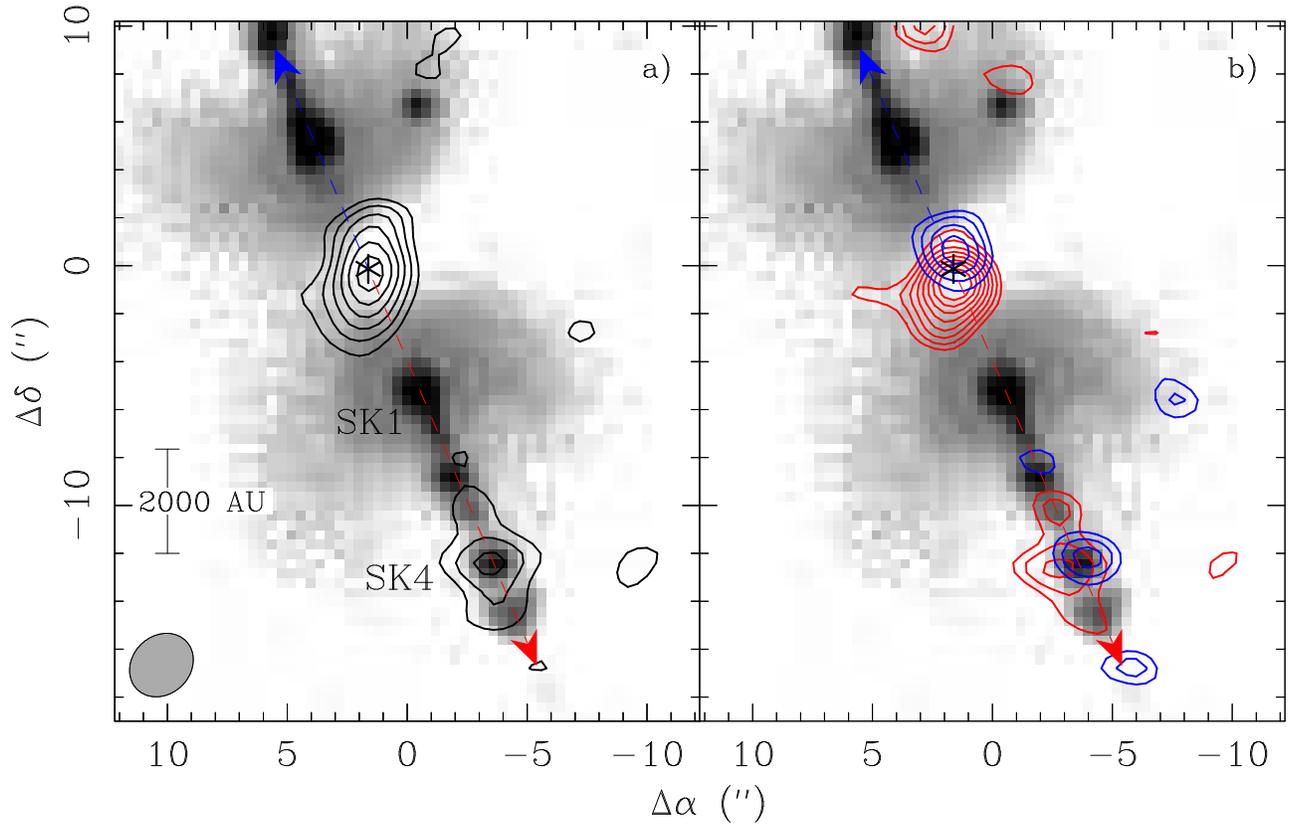

\centering
\putfig{1}{270}{f8.ps}
\figcaption[]{
SO emission plotted on top of the \H2{} image,
with the asterisk indicating the IRAS source.
\tlabel{a} Total emission integrated over 16 \vkm{} from -7 to 9 \vkm{}.
The contours go from 
3 to 13 $\sigma$ with a step of 2 $\sigma$, where $\sigma = 0.35$ \Jybk{}.
\tlabel{b} Redshifted (red contours) and blueshifted (blue contours) 
emission, integrated from -7 to -2 \vkm{} and from 2 to 9 \vkm{},
respectively.
The contours go from 
3 to 10 $\sigma$ with a step of 1 $\sigma$, where $\sigma = 0.23$ \Jybk{}.
\label{fig:SO}
}
\end{figure}

\begin{figure}[!hbp]
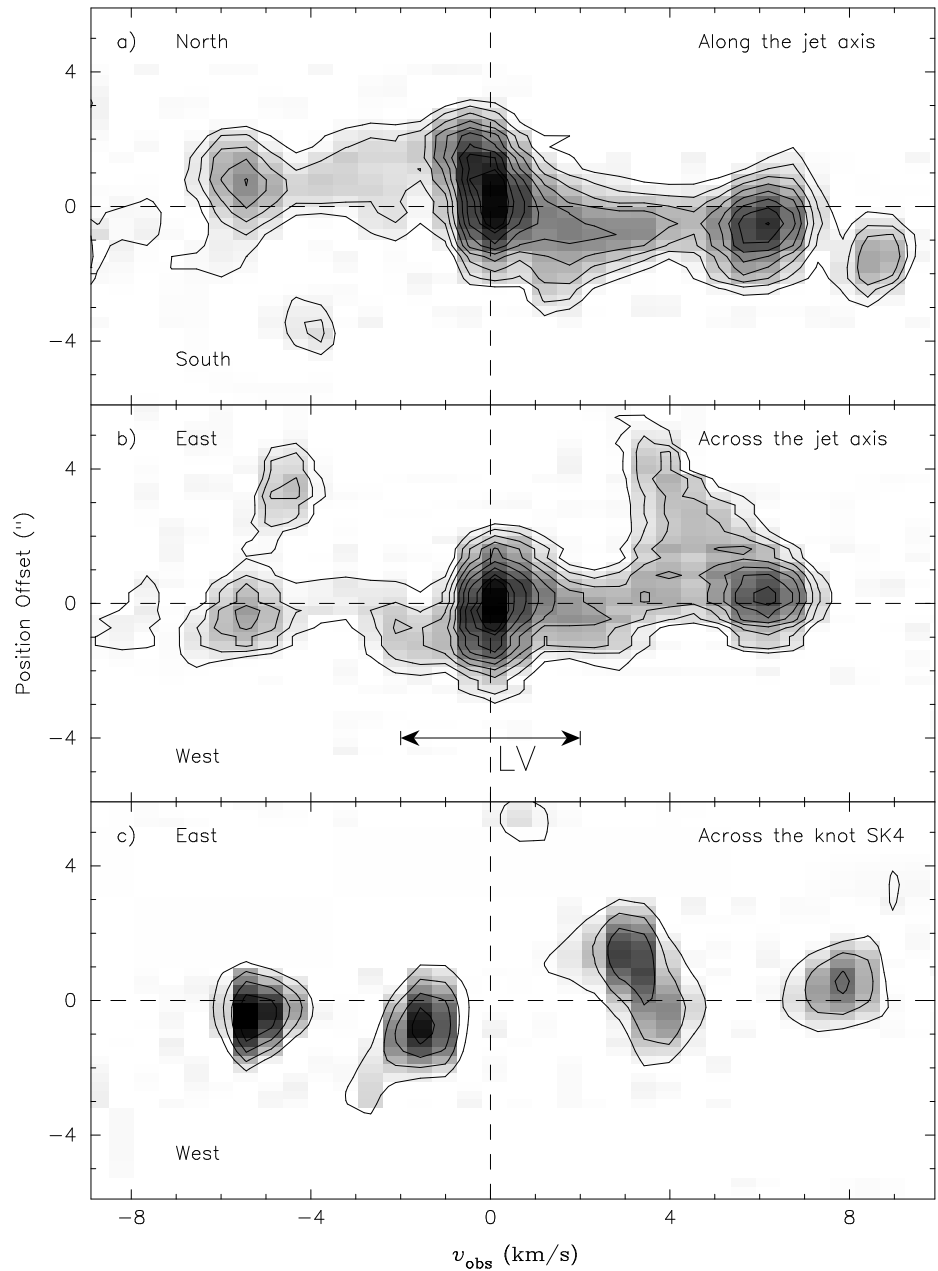

\centering
\putfig{1}{270}{f9.ps}
\figcaption[]{PV diagrams of SO emission. \tlabel{a} A cut along the jet
axis for the emission around the source.
\tlabel{b} A cut across the jet axis for the emission around the source.
LV denotes low velocity.
\tlabel{c} A cut across the jet axis for the emission around the knot SK4.
\label{fig:pvSO}
}
\end{figure}

\begin{figure}[!hbp]
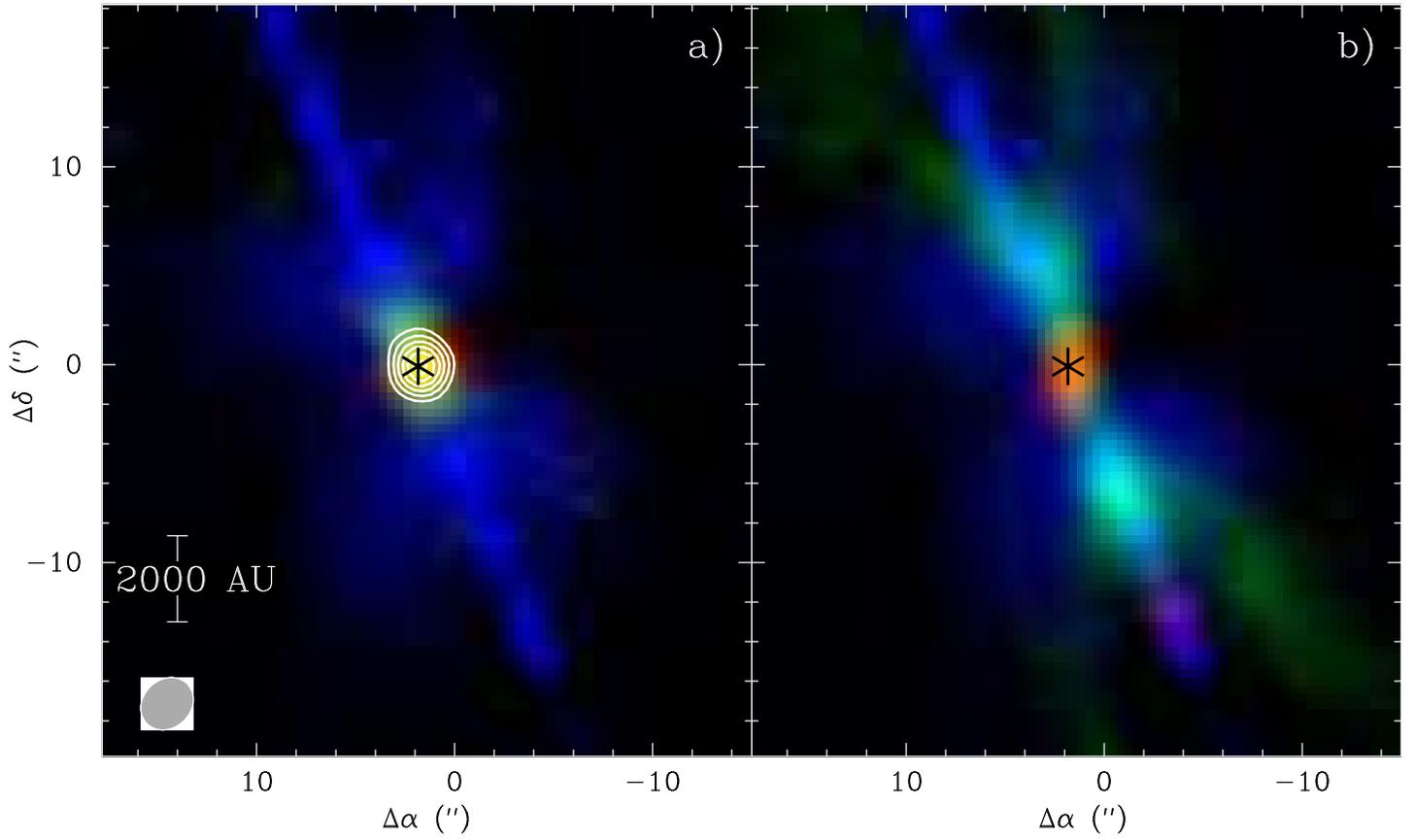

\centering
\putfig{1.3}{270}{f10.ps}
\figcaption[]{
\tlabel{a} Composite image of continuum (white contours), \cCO{} 
(red), \bCO{} (green), and \H2{} (blue) emission.
\tlabel{b} Composite image of SO (red), \aCO{} (green), and \H2{} 
(blue) emission.
\label{fig:RGB}
}
\end{figure}

\begin{figure}[!hbp]
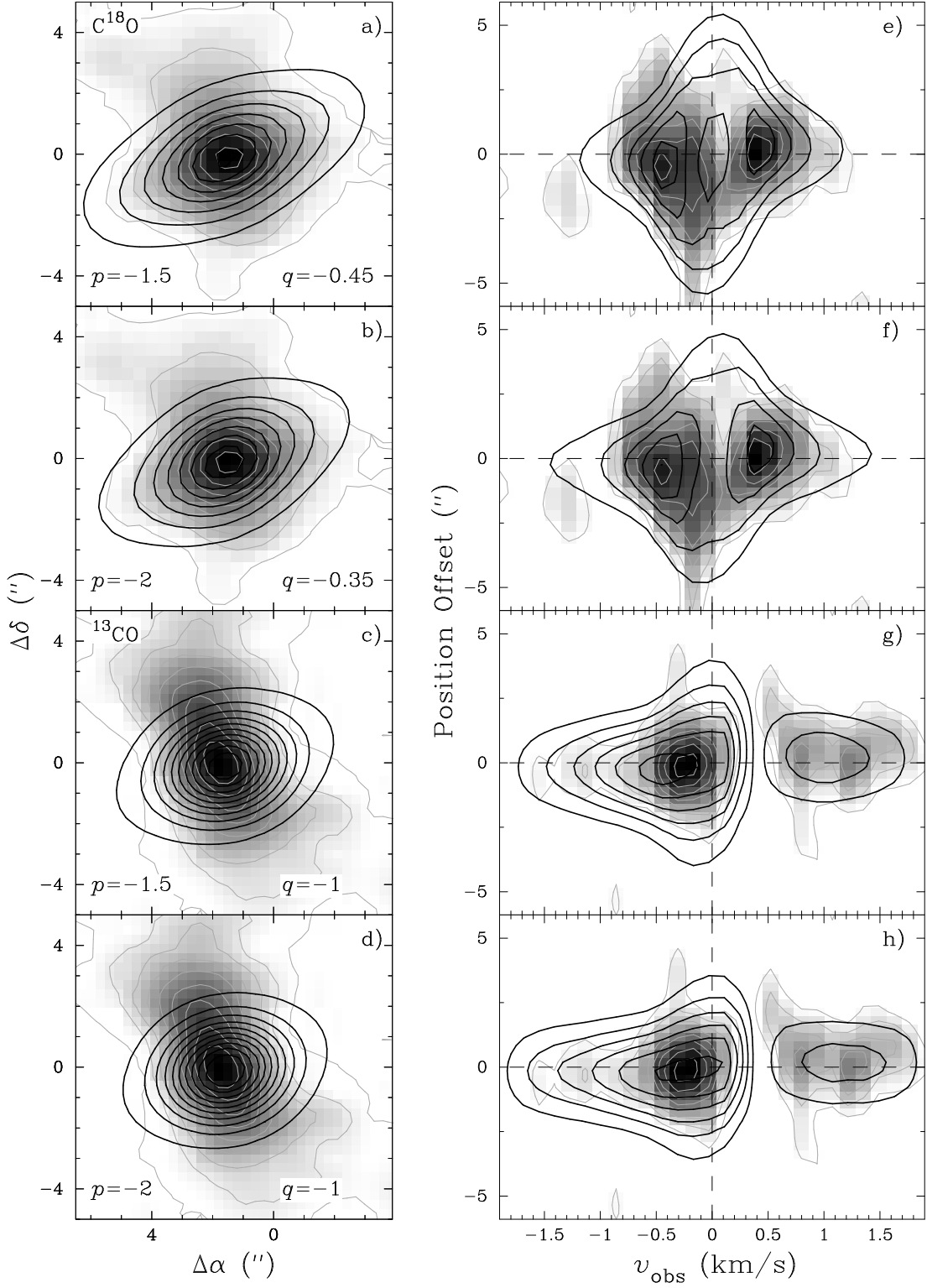

\centering
\putfig{0.85}{0}{f11.ps}
\figcaption[]{
Modeling of envelope in \cCO{} ({\it a,b,e,f})
and \bCO{} ({\it c,d,g,h}).
Gray contours with image are the observations, black contours are derived
from our simple model. \tlabel{a}-\tlabel{d} shows the 
total emission, while \tlabel{e}-\tlabel{h} shows the PV diagrams cut along
the major axis of the flattened envelope.
\label{fig:fpvenv}
}
\end{figure}


\begin{thebibliography}{}
\bibitem[Andre et al.(2000)]{Andre2000} Andre, P., Ward-Thompson, 
D., \& Barsony, M.\ 2000, Protostars and Planets IV, 59  
\bibitem[Beckwith et al.(1990)]{Beckwith1990} Beckwith, S.~V.~W., 
Sargent, A.~I., Chini, R.~S., \& Guesten, R.\ 1990, \aj, 99, 924 
\bibitem[Chapman et al.(2002)]{Chapman2002} Chapman, N.~L., Mundy, 
L.~G., Lee, C.-F., \& White, S.~M.\ 2002, Bulletin of the American 
Astronomical Society, 34, 1133 
\bibitem[Chini et al.(1997)]{Chini1997} Chini, R., Reipurth, B., 
Sievers, A., Ward-Thompson, D., Haslam, C.~G.~T., Kreysa, E., \& Lemke, R.\ 
1997, \aap, 325, 542 
\bibitem[Claussen et al.(1998)]{Claussen1998}Claussen, M.J., Marvel., K.B., 
  Wootten, A., Wilking, B.A. 1998, ApJL, 507, L79
\bibitem[Codella et al.(2005)]{Codella2005} Codella, C., Bachiller, 
R., Benedettini, M., Caselli, P., Viti, S., \& Wakelam, V.\ 2005, \mnras, 
361, 244 
\bibitem[Davis et al.(2000)]{Davis2000} Davis, C.~J., Berndsen, 
A., Smith, M.~D., Chrysostomou, A., \& Hobson, J.\ 2000, \mnras, 314, 241 
\bibitem[Evans(1999)]{Evans1999} Evans, N.~J.\ 1999, \araa, 37, 311 
\bibitem[Frerking et al.(1982)]{Frerking1982} Frerking, M.~A., 
 Langer, W.~D., \& Wilson, R.~W.\ 1982, \apj, 262, 590 
\bibitem[Galli \& Shu(1993)]{Galli1993} Galli, D., \& Shu, F.~H.\ 
 1993, \apj, 417, 243 
\bibitem[Galv{\' a}n-Madrid et al.(2004)]{Galvan2004} Galv{\' 
a}n-Madrid, R., Avila, R., \& Rodr{\'{\i}}guez, L.~F.\ 2004, Revista 
Mexicana de Astronomia y Astrofisica, 40, 31 
\bibitem[Gibb et al.(2004)]{Gibb2004} Gibb, A.~G., Richer, 
J.~S., Chandler, C.~J., \& Davis, C.~J.\ 2004, \apj, 603, 198 
\bibitem[Gueth et al.(1997)]{Gueth1997} Gueth, F., Guilloteau, S., Dutrey, 
 A., \& Bachiller, R.\ 1997, \aap, 323, 943 
\bibitem[Gueth \& Guilloteau(1999)]{Gueth1999} Gueth, F. \& 
  Guilloteau, S. 1999, \aap, 343, 571 
\bibitem[Hartmann et al.(1996)]{Hartmann1996} Hartmann, L., Calvet, 
N., \& Boss, A.\ 1996, \apj, 464, 387
\bibitem[Hayashi et al.(1993)]{Hayashi1993} Hayashi, M., Ohashi, 
N., \& Miyama, S.~M.\ 1993, \apjl, 418, L71 
\bibitem[Hogerheijde(2001)]{Hogerheijde2001} Hogerheijde, M.~R.\ 2001, 
\apj, 553, 618
\bibitem[Lee et al.(2000)]{Lee2000} Lee, C.-F., Mundy, L.G., Reipurth, B.,
  Ostriker, E.C., \& Stone, J.M. 2000, \apj, 542, 925
\bibitem[Lee et al.(2001)]{Lee2001} Lee, 
 C.-F., Stone, J.~M., Ostriker, E.~C., \& Mundy, L.~G.\ 2001, \apj, 557, 429  
\bibitem[Lin et al.(1994)]{Lin1994} Lin, D.~N.~C., Hayashi, M., 
Bell, K.~R., \& Ohashi, N.\ 1994, \apj, 435, 821
\bibitem[McCaughrean et al.(2002)]{McCaughrean2002} McCaughrean, M., 
Zinnecker, H., Andersen, M., Meeus, G., \& Lodieu, N.\ 2002, The Messenger, 
109, 28 
\bibitem[Momose et al.(1998)]{Momose1998} Momose, M., Ohashi, N., 
 Kawabe, R., Nakano, T., \& Hayashi, M.\ 1998, \apj, 504, 314 
\bibitem[Nakamura(2000)]{Nakamura2000} Nakamura, F.\ 2000, \apj, 
543, 291
\bibitem[Ohashi et al.(1997)]{Ohashi1997} Ohashi, N., Hayashi, M., 
Ho, P.~T.~P., \& Momose, M.\ 1997, \apj, 475, 211 
\bibitem[Shirley et al.(2000)]{Shirley2000} 
 Shirley, Y.~L., Evans, N.~J., Rawlings, J.~M.~C., \& Gregersen, E.~M.\ 
 2000, ApJS, 131, 249 
\bibitem[Shu(1977)]{Shu1977} Shu, F.~H.\ 1977, \apj, 214, 488 
\bibitem[Shu, Adams, \& Lizano(1987)]{Shu1987} Shu, F.~H., 
 Adams, F.~C., \& Lizano, S.\ 1987, \araa, 25, 23 
\bibitem[Shu et al.(2000)]{Shu2000} Shu, F.H.,  Najita, J., Shang, H., 
  \& Li, Z. -Y. 2000, in Protostars and Planets IV, ed. V. 
  Mannings, A. P. Boss \& S. S. Russell 
  (Tucson: University of Arizona Press), in press
\bibitem[Wiseman et al.(2001)]{Wiseman2001} Wiseman, J., Wootten, 
A., Zinnecker, H., \& McCaughrean, M.\ 2001, \apjl, 550, L87 
\bibitem[Zinnecker, McCaughrean, \& Rayner(1998)]{Zinnecker1998} 
  Zinnecker, H., McCaughrean, M. J. \& Rayner, J. T. 1998, \nat, 394, 862 
\bibitem[Zinnecker et al.(1992)]{Zinnecker1992} Zinnecker, H. , 
  Bastien, P. , Arcoragi, J. -P.  \& Yorke, H. W. 1992, \aap, 265, 726 
\end{thebibliography}
\end{document}